\documentclass[12pt,preprint]{aastex}

\shorttitle{Bounding the mass of the graviton}
\shortauthors{Jones}

\begin{document}

\title{Bounding the mass of the graviton using eccentric binaries}

\author{D. I. Jones\altaffilmark{1}}


\altaffiltext{1}{Center for Gravitational Wave Physics, The
Pennsylvania State University, State College, Pennsylvania,
16802-6300.  Present address: School of Mathematics, University of
Southampton, Southampton SO17 1BJ, UK; dij@maths.soton.ac.uk}


\date{\today}

\clearpage 

\begin{abstract}

We describe a method by which gravitational wave observations of
eccentric binary systems could be used to test General Relativity's
prediction that gravitational waves are dispersionless.  We present
our results in terms of the graviton having a non-zero rest mass, or
equivalently a non-infinite Compton wavelength.  We make a rough
estimate of the bounds that might be obtained following gravitational
wave detections by the space-based LISA interferometer.  The bounds we
find are comparable to those obtainable from a method proposed by
Will, and several orders of magnitude stronger than other dynamic
(i.e. gravitational wave based) tests that have been proposed.  The
method described here has the advantage over those proposed previously
of being simple to apply, as it does not require the inspiral to be in
the strong field regime nor correlation with electromagnetic signals.
We compare our results with those obtained from static
(i.e. non-gravitational wave based) tests.

\end{abstract}

\keywords{gravitation --- gravitational waves --- relativity}

\clearpage

\section{Introduction}
\label{sect:i}

In Einstein's theory of General Relativity, linearization of the field
equations shows that small perturbations of the metric obey a wave
equation (Misner, Thorne \& Wheeler 1973).  These small disturbances,
referred to as gravitational waves, travel at the speed of light.
However, some other gravity theories predict a dispersive propagation
(see \citet{wy04} for references).  The most commonly considered form
of dispersion supposes that the waves obey a Klein--Gordan type
equation:
\begin{equation}
\label{eq:kg}
\left[ \frac{1}{c^2}\frac{\partial^2}{\partial t^2} - \nabla^2 
+ \left(\frac{m_{\rm g}c}{h}\right)^2
\right] \psi = 0.
\end{equation}
Physically, the dispersive term is ascribed to the quantum of
gravitation having a non-zero rest mass $m_{\rm g}$, or equivalently a
non-infinite Compton wavelength $\lambda_{\rm g} = h/m_{\rm g}c$.  The
group velocity of propagation for a wave of frequency $f_{\rm gw}$ is
then
\begin{equation}
\label{eq:vg}
v_{\rm g} \approx c \left[ 1 - \frac{1}{2} 
  \left(\frac{c}{f_{\rm gw}\lambda_{\rm g}}\right)^2\right],
\end{equation}
valid for $\lambda_{\rm gw} \ll \lambda_{g}$; only in the infinite
frequency limit is General Relativity recovered, with waves traveling
at the speed of light \citep{will98}.

Over the past few decades a number of different \emph{dynamic} tests
of this dispersive hypothesis have been described, i.e. tests making
use of direct observations of gravitational waves or their radiation
reaction effects \citep{chl03,fs02,lh00,will98,wy04}.

In this paper we add another method to this list; we consider
gravitational radiation from \emph{eccentric} binary systems.  Such
binaries emit gravitational radiation at (infinitely many) harmonics
of the orbital frequency \citep{pm63}.  Our idea lies simply in
measuring the phase of arrival of these harmonics.  Dispersion of the
form described by equation (\ref{eq:vg}) would be signaled by the
higher harmonics arriving slightly earlier than the lower harmonics,
as compared to the General Relativistic waveform.  We present a rough
estimate of the bounds that might be obtained, deferring a more
accurate calculation to a future study (Barack \& Jones, in
preparation).

The plan of this paper is as follows.  In \S \ref{sect:dotb} we derive
formulae to make a simple estimate of the bounds that might be
obtained using our method.  In \S \ref{sect:r} we estimate bounds
obtainable on $\lambda_{\rm g}$ for LISA observations of two sorts of
binary systems.  Finally in \S \ref{sect:c} we summarize our findings
and compare with those of other authors.

\section{Derivation of the bound}
\label{sect:dotb}

\subsection{General formula}

To derive a rigorous estimate of the bound one should add the graviton
mass to the list of unknown source parameters to be extracted from the
measured signal, as was done by Will in the case of circular orbits
\citep{will98}.  The $m_{\rm g}$--dependent waveform can then be
computed, allowing calculation of the Fisher information matrix
$\Gamma_{ab}$, which could then be inverted, the $\Gamma_{m_{\rm g}
m_{\rm g}}^{-1}$ component, evaluated at $m_{\rm g} = 0$, giving the
best bound obtainable \citep{will98}.  For the case of eccentric
binaries such a calculation is not easy, and so in this paper we make
a preliminary estimate of the possible bounds, without going to the
trouble of calculating $\Gamma_{\rm ab}$.

We will begin by deriving a general formula for estimating the bound
on $\lambda_{\rm g}$ that could be obtained from a system which
produces gravitational waves at two different frequencies, say $f_{\rm
gw,1}$ and $f_{\rm gw,2}$.  The two gravitational waves will travel
with (different) speeds $v_{\rm g,1}$ and $v_{\rm g,2}$, and so their
journey times to the detector a distance $d$ away will differ by a
time interval $\Delta t$ given by
\begin{equation}
\Delta t = \frac{dc}{2\lambda_{\rm g}^2} 
\left[ \frac{1}{f_{\rm gw,1}^2} - \frac{1}{f_{\rm gw,2}^2} \right]. 
\end{equation}
Multiplying this by $2\pi f_{\rm gw}$, where $f_{\rm gw}$ is a
characteristic frequency in the problem, gives the accumulated
difference in phase of arrival of the two signals caused by the
dispersion, measured in terms of radians of phase of $f_{\rm gw}$:
\begin{equation}
\Delta \Phi_{\rm dispersion} = \frac{\pi c}{\lambda_{\rm g}^2} d \, 
f_{\rm gw}
\left[ \frac{1}{f_{\rm gw,1}^2} - \frac{1}{f_{\rm gw,2}^2} \right].
\end{equation}
This is to be compared with the accuracy with which the phase of
arrival of the waves can be extracted from the noisy gravitational
wave data stream.  In the high signal to noise ratio regime the error
in extracting the phase of a continuous signal can be written as
\begin{equation}
\label{eq:DeltaPhi}
\Delta \Phi_{\rm error} \approx \frac{\alpha}{2\rho},
\end{equation}
where we follow the notation of \cite{chl03}.  In this formula $\rho$
is the signal to noise ratio of the measurement and $\alpha$ is a
dimensionless factor that depends upon how many unknown parameters
(including the phase) need be extracted from the signal.

The lower bound that can be placed on $\lambda_{\rm g}$ comes from
equating $\Delta \Phi_{\rm dispersion}$ and $\Delta \Phi_{\rm error}$
to give:
\begin{equation}
\label{eq:lambdageneral}
\lambda_{\rm g}^2 > 2\pi c \frac{\rho d}{\alpha} 
f_{\rm gw} 
\left[ \frac{1}{f_{\rm gw,1}^2} - \frac{1}{f_{\rm gw,2}^2} \right].
\end{equation}
This shows that the best bounds will come from high mass (i.e. high
$\rho d$), high eccentricity, low orbital frequency systems.

\subsection{Application to eccentric binary systems}

We will now apply this method of estimation to eccentric binary
systems.  In general many more than two harmonics will contribute
significantly to $\rho$, so equation (\ref{eq:lambdageneral}) is not
directly applicable.  In order to take advantage of this spread we
will make the following identifications.  We will set $\rho$ equal to
the total signal to noise of the observation.  To identify appropriate
frequencies, consider a plot of the signal to noise of the n-th
harmonic, $\rho_{n}$, verses gravitational wave frequency $n f_{\rm
orbit}$.  We will set $f_{\rm gw}$ to be the frequency at which this
curve peaks, and $f_{\rm gw,1}, f_{\rm gw,2}$ as the frequencies
corresponding to the lower and upper full-width-at-half-maximum.  In
reality only discrete harmonic frequencies exist, but for the purpose
of defining $f_1$, $f_2$ and $f_{\rm gw}$, we will treat the curve as
continuous, interpolating to find the necessary frequencies.  (A
formalism using only discrete frequencies would have introduced
spurious step-wise changes in our bounds on $\lambda_{\rm g}$ as a
function of eccentricity).

Identification of a suitable $\alpha$ value, which quantifies the
error in phase measurement, is more problematic.  \cite{chl03}
consider errors in measuring the phase of a single monochromatic
signal of known sky location; they find that $\alpha < 3$ for a large
fraction of the possible binary orientations.  \cite{bc04} examine
extreme mass ratio inspirals.  They find phase measurement errors
which again yield $\alpha \approx 3$ (see the $\Delta(t_0) \nu_0$
parameter of their Table III).  

However, even in the dispersionless case of General Relativity, the
relative phasing of the detected harmonics is non-trivially determined
by the source's sky location and the relative orientation of the
detector and binary system \citep{bc04}.  The phase differences we are
considering here are \emph{additional} delays caused by dispersive
propagation.  Clearly, then, the results of \cite{chl03} and
\cite{bc04} do not directly apply to our problem.  Only a full Fisher
matrix calculation will accurately show how we can disentangle the
phase differences contributed by dispersion, measurement error and
those intrinsic to the binary.  We expect that in those situations
where the system parameters, including its sky location and
orientation relative to LISA, are measured accurately, the
dispersion-induced phase delays will be measured accurately too.  In
the absence of a full Fisher matrix calculation to evaluate the
correct measurement errors we will set $\alpha = 10$, but note that
this is the weakest link in our estimate.

In particular, if the various geometric factors that enter the problem
conspire such that a dispersionless signal from a certain binary is
very similar to the dispersed signal from a binary with slightly
different parameters (e.g. a slightly different sky location), then
the errors in $\Delta \Phi$ could be very much larger than estimated
here.  Also, $\alpha$ will depend upon the type of system being
studied.  It will generally be smaller for systems where information
in addition to the gravitational wave signal is available,
e.g. Galactic binary systems where optical measurements give accurate
sky locations.  Note, however, that $\alpha$ enters the bound on
$\lambda_{\rm g}$ only rather weakly, as $\lambda_{\rm g} \propto
1/\sqrt{\alpha}$, and so we hope that our ignorance of this factor
will not change our qualitative conclusions.

\section{Results}
\label{sect:r}

It is expected that gravitational radiation reaction will result in
most binary systems detectable by ground based interferometers being
nearly perfectly circular \citep{pete64} and so will be unusable for
deriving a bound of the sort described here.  We will therefore
concentrate exclusively on (two sorts of) binaries in the LISA band.
  In equation
(\ref{eq:lambdageneral}) we will set $\alpha = 10$, as discussed
above.  When calculating $\rho$ we will assume an integration time of
one year.  We computed the LISA noise using the Online Sensitivity
Curve
Generator\footnote{http://www.srl.caltech.edu/\~{}shane/sensitivity}
\citep{lars03}, which included a fit to the Galactic white dwarf
background \citep{bh97}.

\subsection{Extra-Galactic extreme mass-ratio binaries}

We consider here the inspiral of a solar-mass type black hole into a
massive one.  These are excellent systems from our point of view, as
they are expected to dominate the LISA inspiral event rate and,
crucially, many will have very large eccentricities \citep{bc04,
getal04}.

To see if such systems can indeed be used to obtain a bound on
$\lambda_{\rm g}$, in Figure \ref{fig1} we plot the
eccentricity-orbital frequency phase space for a $(10^6, 10^1)
M_\odot$ binary at a distance of $1$\,Gpc.  The upper curve describes
the innermost stable orbit (ISO) \citep{will98}; binary systems in
Nature only exist \emph{below} this curve.  The lower curve gives the
minimum eccentricity required for the system to be detectable, with
multiple harmonics contributing significantly to $\rho$.  [Our exact
criterion is to see if $\rho$ exceeds some detection threshold
$\rho_{\rm min}$ when the single strongest harmonic is removed from
the sum.  We have set $\rho_{\rm min} = 15$, as would be reasonable if
computational power does not limit the search \citep{getal04}].  Our
methods are only applicable for systems \emph{above} this curve.  It
follows that we can use binary systems which lie \emph{between} these
two curves to bound $\lambda_{\rm g}$.  Fortunately we see that this
means that binaries in a significant portion of the $e-f$ plane are of
use to us.  To illustrate this, a trajectory of a plausible LISA
source is shown between the two curves, with an eccentricity at the
ISO of about $0.24$.  This system spends about $10$ years between the
two curves.

In Figure \ref{fig2} we show the actual bounds on
$\lambda_{\rm g}$ that could be obtained from observations of extreme
mass ratio systems.  The distance is still fixed at $1$\,Gpc, but now
we fix the orbital frequency at $10^{-3}$\,Hz and leave the
eccentricity as a free parameter.  Results for binary systems with
$M_1 = 10^6 M_\odot$ and several different values of $M_2$ are given,
as indicated.

The following features are of note: (i) Each curve terminates at a
minimum eccentricity below which the system is undetectable and/or
fewer than two harmonics contribute significantly to $\rho$, and at a
maximum eccentricity above which the system is dynamically unstable.
(ii) For a system of given masses, the bound increases slightly
(i.e. becomes stronger) the larger the eccentricity.  (iii) Stronger
bounds are obtained from more massive systems, and can be obtained for
wider ranges of the eccentricity.

\subsection{Stellar mass Galactic binary systems}

LISA will be able to detect gravitational waves from a large number of
low mass Galactic binaries, consisting of white dwarfs and/or neutron
stars \citep{detal96}.  To investigate the suitability of these
systems for bounding $\lambda_{\rm g}$, in Figure \ref{fig3} we plot
the eccentricity--frequency phase space for a Galactic $(1, 1)
M_\odot$ binary at a distance $1$\,kpc.  We set $\rho_{\rm min} = 8$,
although a lower value could be used for electromagnetically studied
binaries \citep{detal96}.  We do not show the ISO curve as for all
plausible eccentricities such a binary would go dynamically unstable
in the much higher LIGO frequency band.  Clearly, binaries in a large
portion of the phase space are of use for bounding $\lambda_{\rm g}$.
However, unlike the case of the extreme mass ratio inspiral, there is
no compelling reason to expect the eccentricities of these systems to
be large.  Many of them will have gone through a period of mass
transfer in the past, which is believed to be an efficient
circularizer.  Nevertheless, as we require merely \emph{one or more of
them} to have a sufficiently large eccentricity, greater than about $5
\times 10^{-3}$, a bound on $\lambda_{\rm g}$ may well be obtained.

In Figure \ref{fig4} we present the bounds on $\lambda_{\rm g}$ that
would be obtained from observations of various equal-mass binaries at
a distance of $1$\,kpc and with an orbital frequency $10^{-3}$\,Hz.
The qualitative form is the same as in Figure \ref{fig2}, except we
terminate the curves at the high eccentricity end at $e = 0.55$ as
such extreme eccentricities seem unlikely.

\section{Comparison with previous methods and summary}
\label{sect:c}

In Table \ref{table:lambdag_dynamic} we collect together reported and
proposed dynamic bounds on $\lambda_{\rm g}$ that have appeared in the
literature, and add two proposed bounds from this work.  As is clear
from perusal of the Table and Figures \ref{fig2} and \ref{fig4}, the
bounds presented here for low mass Galactic systems are comparable to
those of \cite{chl03}, while our bounds from extreme mass ratio
inspirals are comparable to those of \cite{wy04} for massive black
hole coalescence.  It should be remembered that our numbers can only
be regarded as estimates, particularly given our rough guess as to the
accuracy with which dispersion-induced phase delays can be measured.
However, even if $\alpha$, the parameter which quantifies this error,
were four orders of magnitude larger than assumed here, our results
for extreme mass ratio inspirals would still beat both solar system
and Galactic low mass binary tests.

The method presented here has several advantages over other methods.
The analysis of \cite{chl03} requires knowledge of the initial
relative phases of the X-ray and gravitational wave signals from an
accreting white dwarf system; it is not clear if the accretion process
will be sufficiently well understood to allow this.  Less
problematically, the method of \cite{will98} requires knowledge of the
phasing of the binary inspiral waveform in the strongly chirping
regime, as it is this frequency variation that allows the dispersion
test.  In contrast, the method described here is very simple,
requiring only that multiple harmonics can be detected.  It is not
necessary for the binary to be chirping significantly, and correlation
with other (i.e. non-gravitational) radiation is not required.

Returning to equation (\ref{eq:kg}), in the static regime the solution
is of the form of a Yukawa-type potential, i.e. a Newtonian one
suppressed by an exponential $\exp{(-r/\lambda_{\rm g})}$.  This
offers the possibility of bounding $\lambda_{\rm g}$ by looking for
departures from Newtonian gravity in the non-radiative regime.  Such
results are given in Table \ref{table:lambdag_static}.  \cite{tetal88}
used planetary ephemeris data to obtain their bound, while \cite{gn74}
cited evidence of gravitational binding of galaxy clusters, suggesting
that the exponential suppression is not important over length scales
of the order of a Mpc.  The bounds that could be obtained by using the
methods described in this paper would be better than the solar system
bounds by around $5$ orders of magnitude.  However, they are weaker
than those from galaxy clusters by $3$ orders of magnitude.

Therefore, if equation (\ref{eq:kg}) is the correct linearization of
the true theory of gravity, and if galaxy clusters are indeed
gravitationally bound, the bounds on $\lambda_{g}$ from the dynamic
sector are much weaker than those from the static sector.  However,
the possibility remains that equation (\ref{eq:kg}) is not the correct
linearization, the static potential is not suppressed, but the wave
propagation is nevertheless dispersive, i.e. equation (\ref{eq:vg})
holds but is not derived from an equation of the form of equation
(\ref{eq:kg}).  The only way of settling this is to use the methods
proposed in the dynamic regime.  It could even be the case that
neither equations (\ref{eq:kg}) nor (\ref{eq:vg}) are correct, but
that gravitational waves have some other form of dispersion.  The
method considered here (or any of the methods referred to in Table
\ref{table:lambdag_dynamic}) could be used to identify this.

To sum up, the estimates in this paper indicate that our method will
give bounds on $\lambda_{\rm g}$ which are stronger than most other
dynamic bounds, being rivaled only by those of \cite{wy04}. 

Having used simple estimates to establish the competitiveness of the
method presented here with other dynamic tests, we are currently
working to improve the accuracy of our calculation by using the Fisher
information matrix to calculate the bound (rather than the methods of
\S \ref{sect:dotb}; Barack \& Jones, in preparation).  We also aim to
extend the scope of the investigation by considering the full range of
anticipated gravitational wave sources for both ground and space-based
detectors.

\acknowledgments

It is a pleasure to thank Leor Barack, Shane Larson, Ben Owen, Steinn
Sigurdsson and Nico Yunes for useful discussions during this
investigation, and the anonymous referee for providing comments which
improved the manuscript.  The Center for Gravitational Wave Physics is
supported by the National Science Foundation under cooperative
agreement PHY 01-14375.

\clearpage

\begin{deluxetable}{lll}
\tablecaption{Actual/proposed bounds on $\lambda_{\rm g}$ from the
dynamic sector\label{table:lambdag_dynamic}} 
\tablewidth{0pt}
\tablehead{ \colhead{Reference} & \colhead{Binary system} &
\colhead{$\lambda_{\rm g}$ bound} \\ & & (m)} 
\startdata 
1 & Radio pulsars & $1.6 \times 10^{13}$ \\
2 & 4U1820-30 & $1 \times 10^{16}$ \\ 
3 & $e=0.1$, $(0.5, 0.5) M_\odot$ & $1 \times 10^{16}$ \\ 
2 & Ideal low mass binary & $1 \times 10^{17}$ \\ 
4 & ($10^4$--$10^4) M_\odot$ & $4 \times 10^{18}$ \\ 
3 & $e=0.3$, $(10^6, 10^2) M_\odot$ & $3 \times 10^{19}$\\
4 & ($10^7$--$10^7) M_\odot$ & $5 \times 10^{19}$ 
\enddata
\tablerefs{ (1) \cite{fs02}; (2) \cite{chl03}; (3) This work; (4)
\cite{wy04}.}
\end{deluxetable}

\clearpage

\begin{deluxetable}{lll}
\tablecaption{Actual bounds on $\lambda_{\rm g}$ from the static
sector\label{table:lambdag_static}} 
\tablewidth{0pt} 
\tablehead{
\colhead{Reference} & \colhead{System} & 
\colhead{$\lambda_{\rm g}$ bound}  \\ & & (m)} 
\startdata 
1 & Solar system & $2.8 \times 10^{15}$ \\ 
2 & Galaxy clusters & $\sim 10^{23}$
\enddata
\tablerefs{(1) \cite{tetal88}; (2) \cite{gn74}.}
\end{deluxetable}

\clearpage

\figcaption{The eccentricity--orbital frequency phase space for a $(10^6,
  10) M_\odot$ binary at a distance of $1$\,Gpc. The ISO curve, a
  sample trajectory, and the minimum eccentricity $(\rho_{\rm min} =
  15)$ curve are shown. \label{fig1}}

\figcaption{Bounds on $\lambda_{\rm g}$ obtainable from extreme mass
ratio binaries with $f_{\rm orbit}=10^{-3}$\,Hz, $M_1 = 10^6 M_\odot$,
$\rho_{\rm min} = 15$ at a distance of
$1$\,Gpc.\label{fig2}}

\figcaption{The eccentricity--orbital frequency phase space for a $(1, 1)
M_\odot$ binary at a distance of $1$\,kpc.  A sample trajectory and
the minimum eccentricity curve $(\rho_{\rm min} = 8)$ are shown.
\label{fig3}}

\figcaption{Bounds on $\lambda_{\rm g}$ obtainable from equal low mass
Galactic binaries with $f_{\rm orbit}=10^{-3}$\,Hz, $\rho_{\rm min} =
8$ at a distance of $1$\,kpc.\label{fig4}}

\newpage
\plotone{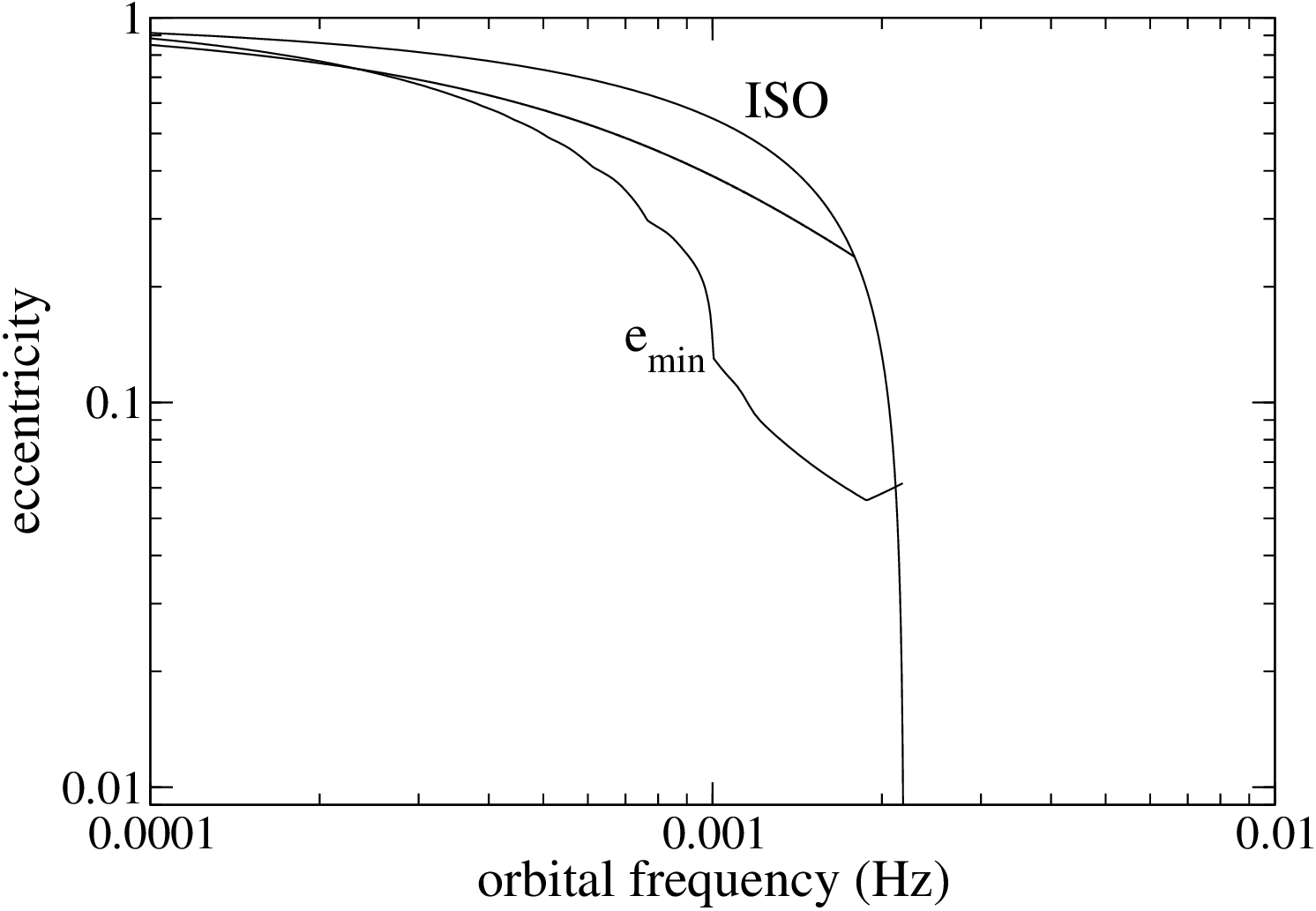}
\newpage
\plotone{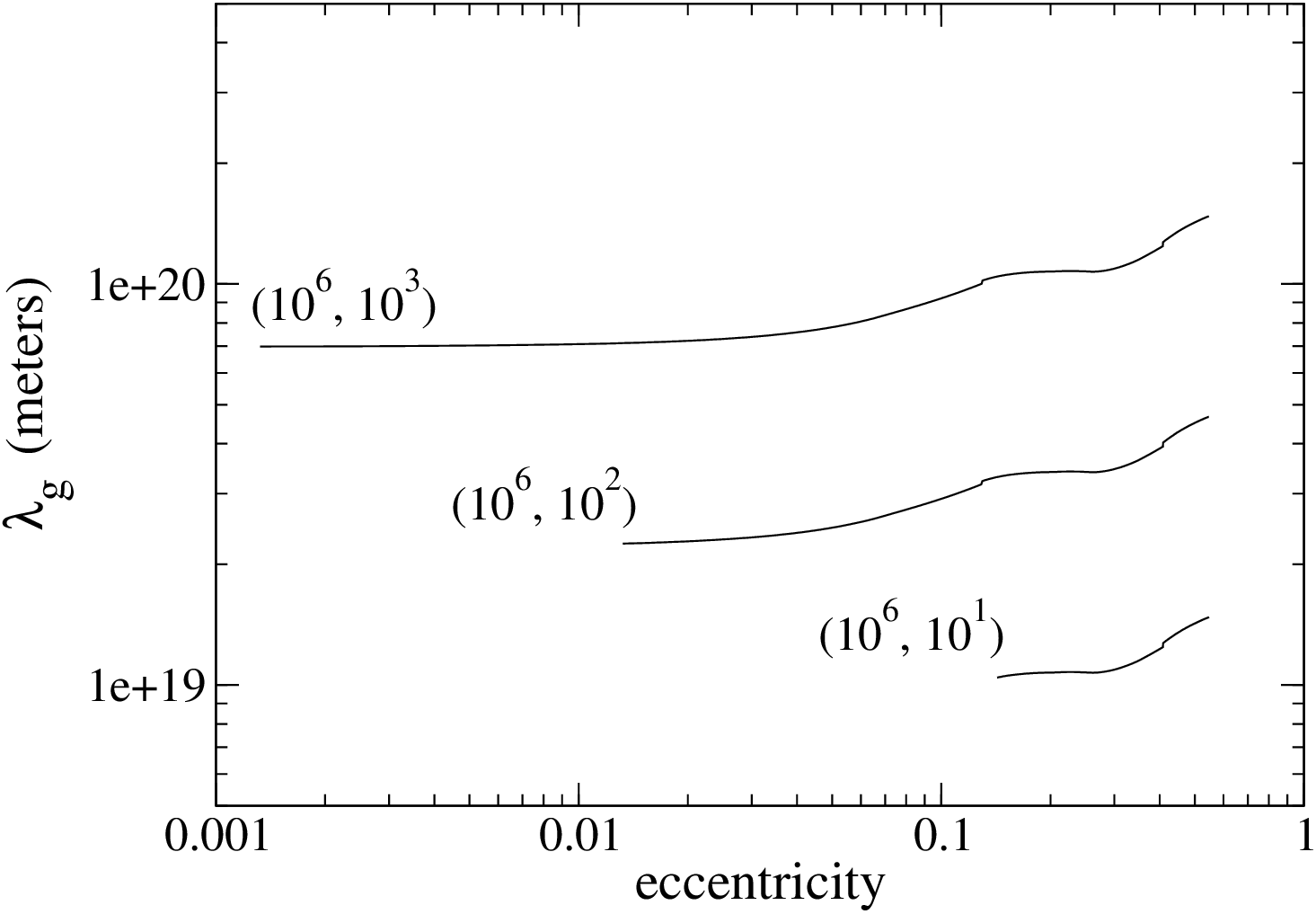}
\newpage
\plotone{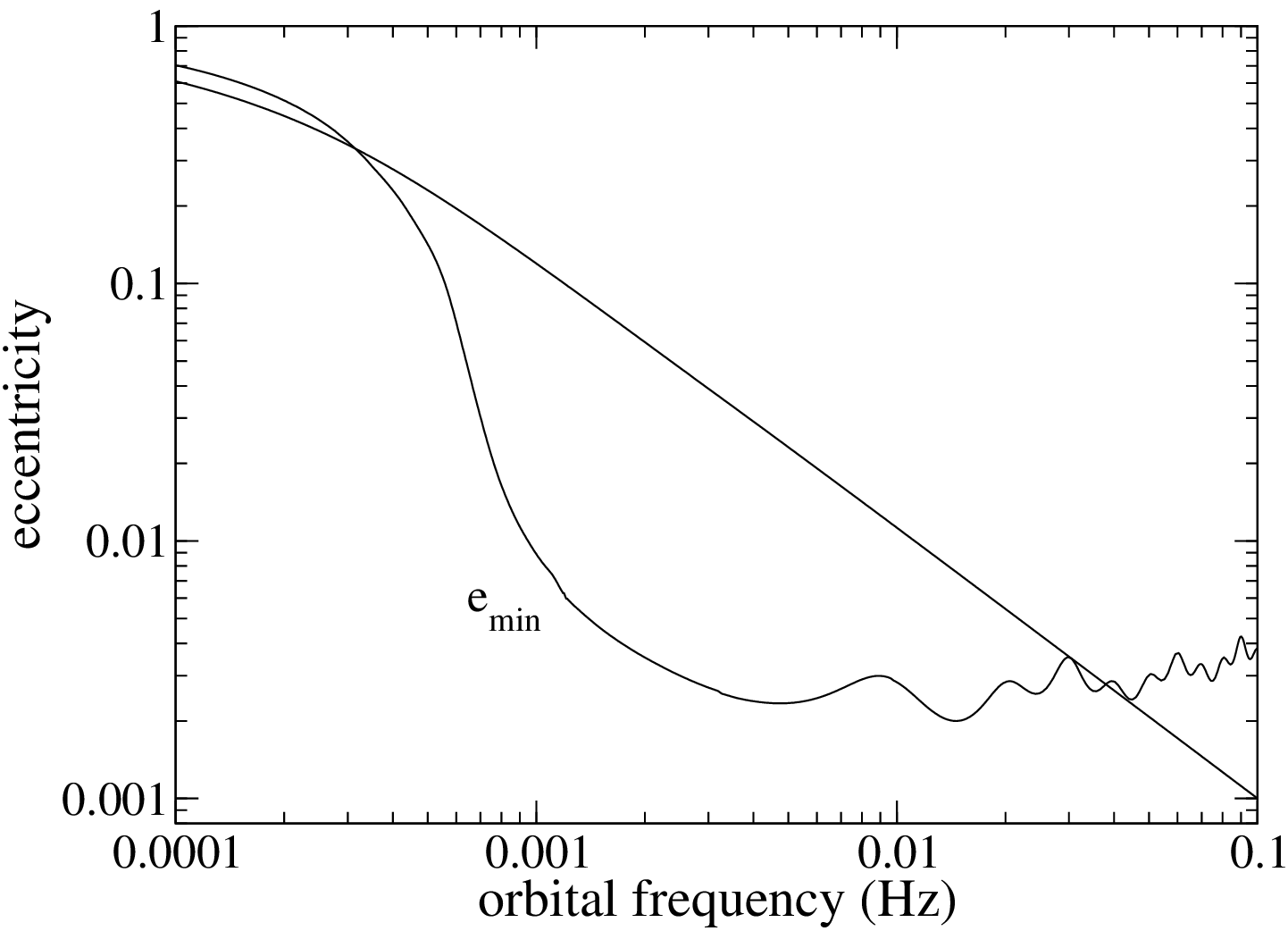}
\newpage
\plotone{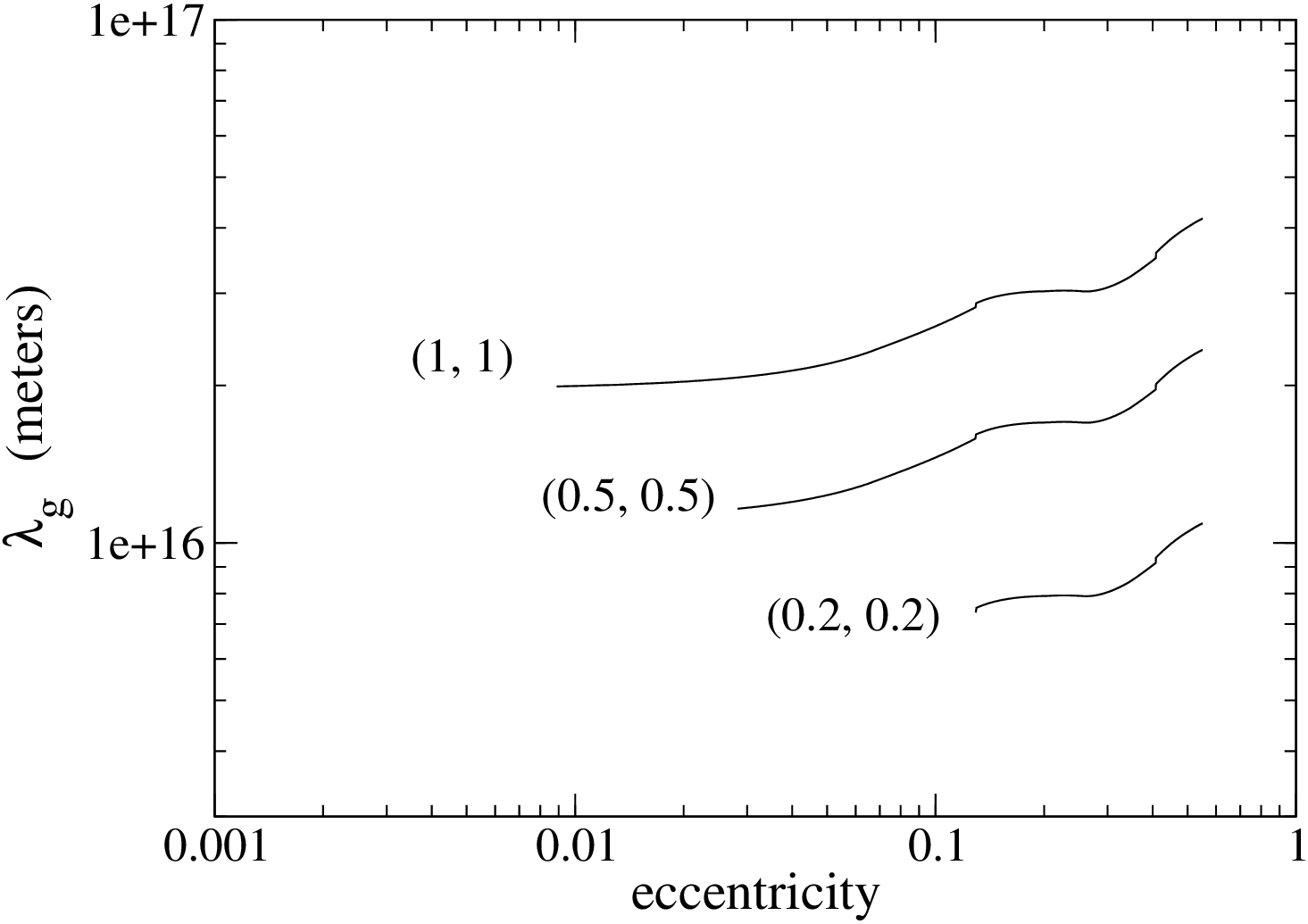}

\end{document}